# THE INTERNET OF THINGS: NEW INTEROPERABILITY, MANAGEMENT AND SECURITY CHALLENGES

Mahmoud Elkhodr, Seyed Shahrestani and Hon Cheung

School of Computing, Engineering and Mathematics, Western Sydney University, Sydney, Australia

## ABSTRACT

*The Internet of Things (IoT) brings connectivity to about every objects found in the physical space. It extends connectivity to everyday objects. From connected fridges, cars and cities, the IoT creates opportunities in numerous domains. However, this increase in connectivity creates many prominent challenges. This paper provides a survey of some of the major issues challenging the widespread adoption of the IoT. Particularly, it focuses on the interoperability, management, security and privacy issues in the IoT. It is concluded that there is a need to develop a multifaceted technology approach to IoT security, management, and privacy.*

## 1. INTRODUCTION

The Internet of Things (IoT) goes beyond the typical computer-based-Internet model to a distributed heterogeneous model of connected things. The state of the art application in the IoT provides IoT services based on utilizing and combining data received from various things. It is a complex system that has the capabilities of sensing information about the environment, capabilities of collecting physiological measurements, and machine operational data, abilities of identifying users, animals, other things, and events in an environment; and the capabilities of processing and communicating these data with other things [1]. Also, it has the capabilities of converting the data into automated instructions that feedback through the communication networks to other things with actuating capabilities. These things will in turn actuate other things, eliminating many human interference roles. Clearly, with such a diverse, complex and heterogeneous model of the IoT, numerous challenges arise. To realize the unique and futuristic characteristics of the IoT, management and security of things, should be well thought-out as one of the fundamental enablers of this technology. There is a need to manage the unprecedented number of things connected to the Internet that generate a large amount of traffics, particularly things with low resources. With billions of things equipped with sensors and actuators entering the digital word using a vast array of technologies, incorporated into devices like lights, electric appliances, home automation systems and a vast number of other integrated machinery devices, transport vehicles, and equipment; management of things become a necessity and cumbersome task.

Towards this aim, this paper reviews some of the significant issues challenging the realization of the IoT with regards to interoperability, management, security, and privacy. In addition, many of these challenges have an associated requirement that needs to be considered. This requirement relates to the nature and capabilities of things. This is because, in general, things in the IoT are characterized as small and lightweight devices that communicate using low-power wireless technologies such as ZigBee. Therefore, the things' resources such as memory, processing power, and battery supply are very limited. This is, in fact, a challenge for the application of many traditional networking, management and security techniques. This challenge adds another dimension to the issues raised above. For instance, it is difficult to achieve security on lightweight things (e.g. parking sensors) compared to traditional computation devices (e.g. a mobile device). This is due to the fact that it is infeasible to apply traditional security cryptographic-based techniques on things with low resources with regard to computation and power resources [2]. Also, things or groups of things are often deployed in remote areas or in areas where accessibility is an issue; which makes changing the things' batteries a difficult task. As such any computation activity that might consume a lot of energies or require heavy computation is considered as unviable. Therefore, addressing these challenges in tandem with the lightweight requirement of things is essential for the successful deployment and advance of the IoT. The remainder of this paper discusses interoperability and its different type in the IoT, WSNs integration issues and the management, security, and privacy issues challenging the IoT.

## 2. INTEROPERABILITY AND INTEGRATION CHALLENGES

Solutions considering the issues associated with information systems' interoperability can be traced back to 1988 [3]; and perhaps even earlier. Wikipedia defines Interoperability as "the ability to make systems and organizations work together" [4]. The IEEE defines interoperability as "the ability of two or more systems or components to exchange information and to use the information that has been exchanged" [5]. Other definitions of interoperability are further tailored according to the particular application's requirements or needs. As a result, different categories of interoperability have been emerging. Technical interoperability [6], Semantic interoperability [7], Syntactic interoperability [8], and Cross-domain interoperability [9] are examples of these categories. All these types of interoperability are needed to support seamless and heterogeneous communications in the IoT. Achieving interoperability is vital for interconnecting multiple things together across different communication networks. It defeats the purpose to have billions of sensors, actuators, tiny and smart devices connected to the Internet if these devices can't actually communicate with each other in a way or another. In fact, for the IoT to flourish, things connecting to the communication networks, which can be heterogeneous, need to be able to communicate with other things or applications.

In traditional computer environments, computer devices are treated equally when connected to the Internet. Their functionalities vary depending on how the users use them. However, in the IoT, each device would be subject to different conditions such as power energy consumption restrictions, communication bandwidth requirements, computation and security capabilities. Additionally, things could be made by various manufacturers that do not necessarily comply with a common standard. Things may also operate using a variety of communication technologies. These technologies do not necessarily connect things to the Internet in the same way a typical computer device usually do. For instance, 6LoWPAN offers interoperability with other wireless 802.15.4 devices as well as with any IP-based devices using a simple bridging device. However, an advanced application layer gateway is required to bridge between ZigBee and non-ZigBee networks [10].

The highly competitive nature of the IoT makes interoperability between things even a more difficult task to achieve. Besides, wireless communication technologies are evolving and changing rapidly. This adds to the complexity of creating interoperable communications in the IoT as well. This inevitability results in heterogeneous devices that cannot communicate with each other which raise many integration issues in the IoT. Service descriptions, common practices, standards and discovery mechanisms [11] are among the many other challenges that also need to be considered before enabling interoperable interactions between things.

## 2.1 Integration issues

The IoT is challenged by fragmented, often unpredictable, deployment of a mixture of devices (e.g. low-power devices with low capabilities versus more capable devices). This is, in fact, constitutes a potential barrier to achieving interoperability in the IoT. Additionally, the present competitive market in the low-power wireless domain, where each organization is trying to push their standard forward is increasing the risk of non-interoperability between IoT devices creating integration issues. While it is essential to provide the end-user with more choices of technologies, certainties should be maintained. This means the IoT requires standards to enable horizontal platforms that are communicable, operable, and programmable across devices, regardless of their make, model, manufacturer, or industry applications [12].

Things with sensing capabilities could be in the form of smart objects such as smart building, smart cars that could incorporate sensing capabilities as part of their designs and functionality, or simply a tiny wireless sensor. The IoT joins this mixture of devices in a heterogeneous integrated information system where things are capable to collaborate, communicate and provide services. Todays' sensors can monitor temperature, ambiance, soil makeup, pollution in the air, noise, presence of objects or movements among other actuating functionalities triggered based on some sensed information. Traditionally, WSNs are built of several "nodes" ranging from a few nodes up to hundreds or even thousands connected to each other. For example, deploying a large number of sensors in a given forest can help detecting fires and alerting the authorities and nearby communities. Thus, integrating things, or group of things, in the form similar to wireless sensors networks (WSNs) in the IoT is expected to play a significant role in the IoT. However, this might not be as easy to achieve as it sounds. Integrating WSNs in the IoT, where sensor nodes dynamically join the Internet, collaborate or communicate with other similar nodes or things open the door to novel challenges. The next section explores the ways in which WSNs could join the Internet as part of the IoT.

### 2.1.1 Network-based Integration

In the Network-based Integration topology, the sensors join the Internet through their network gateway as shown in Figure 1. In the case of a multi-hop mesh wireless topology, the sensors rely on a base node, also known as a sink, which even possesses gateway's capabilities or have a connection with a gateway. The sensors, in this case, are not directly accessible on the Internet. Communications between a sensor of a particular WSN and that of another WSN or/and with other things on the IoT are not going to be direct but via the WSN's base node.

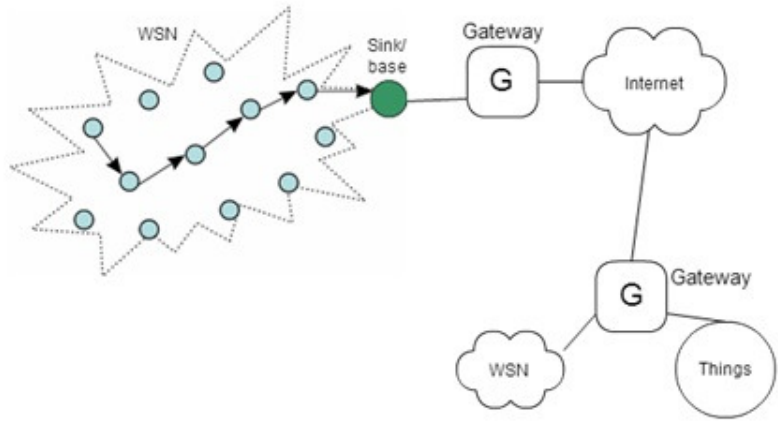

Figure 1- Network Based integration

### 2.1.2 Independent Integration

In the Independent Integration topology, the sensors can connect directly to the Internet independently from their base point. As a result, the interaction between an independent sensor and other things in the IoT can be established without the need to pass by the intermediate node (i.e. the sink node in the WSN). The topology of such a communication network is given in Figure 2. However, giving an IP address to every sensor, for the purpose of connecting to the Internet, may not be the right approach. This is because wireless sensors communications are generally characterised by their low-cost and low-power features with packets exchanged periodically and in small sizes [13]. Therefore, it is quite challenging to provide every IoT device with an IP address to connect to the Internet. This is due to the communication and processing overheads associated with the use of the TCP protocol that challenges the capabilities of small and low-cost sensors [14].

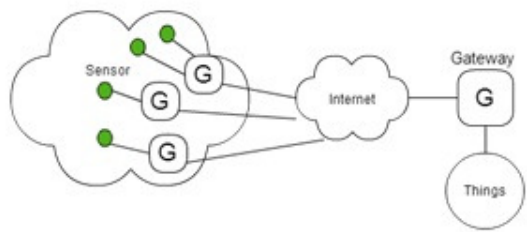

Figure 2- Independent integration

### 2.1.3 Hybrid Integration

Figure 3 shows the "hybrid integration" topology. In this topology, the IoT integrates WSNs using a mixture of the previously introduced topologies.

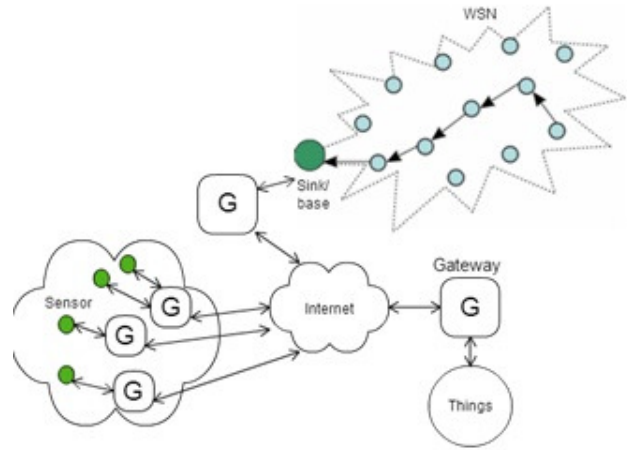

Figure 3- Hybrid integration

The wireless sensor nodes involved in each of the network topologies presented in Figures 1, 2 and 3 may have different characteristics. Their technical requirements vary from one to another. For instance, in the Network-based Integration network, the sensors rely on a base node when connecting to the IoT. The base node possesses more computational, energy and communication resources when compared with a regular sensor node. A base node connects to the Internet in two ways:

(1) Basic: the base point provides basic gateway services such as protocol translation services and routing [15]. It typically forwards the information collected by the WSN's nodes to a server.
(2) Advanced: In addition to its gateway functionality, the base point has the capabilities of computing, and performing some data analysis. This can help reducing redundancy in the network.

Consequently, at least the base node in a WSN requires an IP address to connect to the Internet. In the case of a Network-based Integration network, there is a need to identify uniquely all the sensors on the Internet. Traditionally, the Internet is designed around an address-centric scheme (IP address) as almost all transactions (HTTP, email, etc.) require information about where the data are hosted [16]. Henceforth, a solution based on globally uniquely identifying things should be explored. Dynamic address allocation schemes similar to DHCP and translation schemes similar to DNS need to be exploited as well.

## 2.2 Other Interoperability Challenges

A typical IoT system has been repetitively described in this paper as a system collecting and making use of shared data among things. However for a basic communication to occur, a user or device needs a way to search for things and access the data they produce. This requires an agreement on many fundamental communication issues such as those relating to how things are represented, searched and accessed on the Internet [17]. Additionally, there are associated issues which need to be considered as well. These issues relate to security (authorization, authentication, trust, integrity, validation, etc.), and privacy including privacy of the users and devices. Therefore, to achieve interoperability in the IoT, several other associated challenges need to be considered as well, including the followings:

**Thing Interaction:** As discussed in the previous section, there is more than one option to how things will interact with other things or the users. There exist situations where interactions with individual things are needed. On the other hand, there exist situations where the ability to query and control large groups of things at the same time is also required.

**Virtual Representation of Things**: How things are represented, and described remains an issue unsolved or precisely unstandardized. For instance, do we need to establish a shared schema or ontology for things for greater interoperability? Which attributes should be used to describe things and how flexible and unified this descriptor system should be? Can we find appropriate ways to involve users in connecting things and resolving ambiguities based on their current operation or context? [17].

**Searching, Finding and Accessing Things**: How do we search for things on the Internet? Should we be able to search for things by their unique ID, IP, location, name or/and in combination with other properties? How can we discover, search, locate or track mobile things that may move from one location or network to another? How should things be organized, deployed, managed and secured?

**Syntactic Interoperability between Things:** Recall that syntactic interoperability deals with the packaging and transmission mechanisms for data over a network. Thus, when all the above challenges are addressed, there will still be a need to ensure that data flow is interoperable between the various networks and among a mixture of devices. Translation functionalities in networks or in some devices, gateways or in the form of middleware sitting on the edge of a network are most likely needed.

## 3. MANAGEMENT CHALLENGES

Traditionally, network management solutions are needed to manage network equipment, devices, and services. However, with the IoT, there is a need to manage not only the traditional networked devices and their services, but also an entirely new range of things. The enormous number of things and their diversity create many management requirements. Thus, traditional management functionalities such as remote control, monitoring and maintenance are considered of paramount significance for the operation of things in the IoT. However, these management capabilities need to evolve to cater for the unique characteristics of the IoT. This is because the IoT is of a diverse nature supporting heterogeneous communications and seamless machine to machine interactions. This is in addition to the specific management capabilities required for managing things in the IoT. For example, self-configuration and network reconfiguration are essential management requirements in the IoT. On the other hand, traditionally, network management solutions aimed at providing management information within a minimal response time. However, in some IoT scenarios which might involve lightweight devices, management solutions should provide comprehensive management information with minimal energy use [18].

Nevertheless, the characteristics of data generated in the IoT are distinct from other data in use today [19]. For instance in [20], IoT data have been described as having five distinct characteristics: Heterogeneity, Inaccuracy of sensed data, Scalability and Semantics. We add minimal or constrained as another important feature governing data in the IoT as, generally, things in the IoT have limited computation, communication, and power resources at their disposal. Additionally, in the context of IoT, data management systems must summarize data online from multiple heterogeneous sources while providing storage, logging, and auditing facilities for offline analysis [21, 22]. Therefore, management functionalities are needed to allow managers to perform many maintenance tasks remotely over the Internet and possibly across many heterogeneous interconnected networks. Such management capabilities help in reducing errors and accelerating response time. The ability to turn things on and off, disconnecting things from specific networks, and monitoring the statuses of things are amongst the important tasks that a management system should support. On the other hand, having a management system deployed in an IoT network helps in eliminating travel's and staff training's costs. Also, it helps in accelerating the response to failure events. For example, a management system that supports the remote monitoring, via the Internet, of sensors and other smart devices deployed in remote locations or a busy city is highly beneficial and essential in emergency applications [23]. Such a system allows managers to remotely control, diagnose errors, and troubleshoot IoT devices in real time, reducing costs and accelerating many maintenance tasks.

Furthermore, the magnitude of network connections and data associated with the IoT poses additional challenges in terms of data and service management. These challenges relate to data collection and aggregation, provisioning of services and control as well as monitoring the performance of things. Thus, performance becomes significant in IoT applications that deploy things in remote locations where accessibility is an issue. Performance is also considered important in emergency applications where failure can be catastrophic. Thus, management solutions should provide the capabilities needed to monitor the performance of things and the IoT network as well. Performance statistics relating to response time, availability, up and down time, and others are also highly advantageous. Other performance requirements relate to things' hardware. This is because, providing insights into the health of things, and their networks are an important performance activity. For instance, monitoring and reporting the change in things' state (e.g. the status of an actuator whether it is running or no), the ambience's temperature, hardware's temperature, battery levels, among others, are necessary for the overall management of things in the IoT. Table 1 describes the major management issues challenging the IoT.

Table 1- Management issues

| Configuration Management | <ul><li>How things are setup and by whom?</li><li>Network connectivity</li><li>Self-configuration capability.</li><li>Asynchronous Transaction Support.</li><li>Network reconfiguration.</li></ul> |
|---|---|
| Things' control | Management issues including turning things on and off, disconnecting things from specific networks and connecting to other. To effectively control a thing, a prior knowledge of the thing's status is required. Therefore, "Things' control" complements "monitoring of things." |
| Monitoring | It is essential for the operation and control of things to know the status of things e.g. running, listening, down, sleep mode, etc. Therefore, once things are deployed and in use, there should be a way to monitor their statuses. These are in addition to:<ul><li>Network status monitoring</li><li>Network topology discovery.</li><li>Notification.</li><li>Logging.</li></ul> |
| Things' maintenance: | Detecting the failure of a thing is important, specifically in an IoT network which might involve a larger number of things. A tool or software is required for detecting and addressing things' failure. Other issues relate to the general maintenance tasks of things e.g. software update, patch update, protocols version detections, etc. |
| Things' performance | Monitoring the performance of things is needed so sign of stress can be detected before the occurrence of any failure. This is significant for things that might be deployed in remote locations, and essential in emergency |

| | applications [23], where availability and other QoS parameters are of high importance. |
|---|---|
| Things' security and privacy | There are basic security challenges such as authorization, authentication and access control that need to be addressed. Security bootstrapping mechanisms are also required. Other security issues are associated with things-to-things communications. For instance, if things are to be accessed by applications or software independently from the human users, then there are security measures that need to be enforced to ensure that things are not leaking information and disclosing private information to unauthorized things or used miscellaneously. Things have their users and owners. Thus privacy is vital as well [24]. |
| Energy Management | • Management of energy resources.<br>• Statistics on energy levels, e.g. estimated lifespan. |

## 4. SECURITY CHALLENGES

The growth in the number of connected devices to the communication networks in the IoT translates into increased security risks and poses new challenges to security. A device which connects to the Internet, whether it is a constraint or smart device, inherits the security risks of today's computer devices. Almost all security challenges are found in the IoT. Hence, some fundamental security requirements in the IoT such as authorization, authentication, confidentiality, trust, and data security need to be considered.

Therefore, things should be securely connected to their designated network(s), firmly controlled and accessed by authorized entities. Data generated by things need to be collected, analysed, stored, dispatched and always presented in a secure manner. Nevertheless, there are security risks associated with things-to-things communications as well. This is in addition to the risks relating to things-to-person communications. For instance, if things are to be accessed by things independently from the human users, then there are security measures that need to be enforced. These security measures are necessary to ensure that things are accessed only by authorized entities in a secure manner. Also, they need to ensure that things are not leaking information or disclosing private information to unauthorized things and users, or used miscellaneously.

### 4.1. The Inherited Security Challenges in the IoT

The IoT can be regarded as the Internet 2.0 or the future Internet. Thus, the IoT is not another form of communications or networks running in parallel with what we now today as the Internet; but indeed an expansion of it. Therefore, the IoT inherits today's Internet security issues and poses some new ones as well. Figure 4 shows the security issues inherited and challenging the IoT. These security issues are discussed in the followings sub-sections.

#### 4.1.1. End-to-End Security

Cisco defines end-to-end security as an absolute requirement for secure communications [25]. It is the process of protecting the communications and data exchanged between both ends of the communication without being read, eavesdropped, intercepted, modified, or tampered. In the IoT, end-to-end security remains an open challenge for many IoT devices and applications. The nature of the IoT with its heterogeneous architecture and devices involve the sharing of information and collaboration between things across many networks. This poses serious challenges to the end-to-end security. When devices have different characteristics and operate using a variety of communication technologies (802.11 vs. 802.15.4), establishing secure sessions and secure communications, become a very complex task to achieve.

Additionally, not all devices in the IoT are equal. Currently, computers, smartphones, and other computerized devices connect to the Internet via HTTP, SMTP and the like for most of their activities. As such TLS and IPsec protocols are usually used to negotiate dynamically the session keys, and to provide the required security functions. However, some of the devices in the IoT do not possess the ability to run TLS and IPsec protocols due to their limited computation and power capabilities. Additionally, some embedded devices in the IoT have limited connectivity as such they may not necessarily use HTTP or even IP for the communications (e.g. a sensor in a WSN).

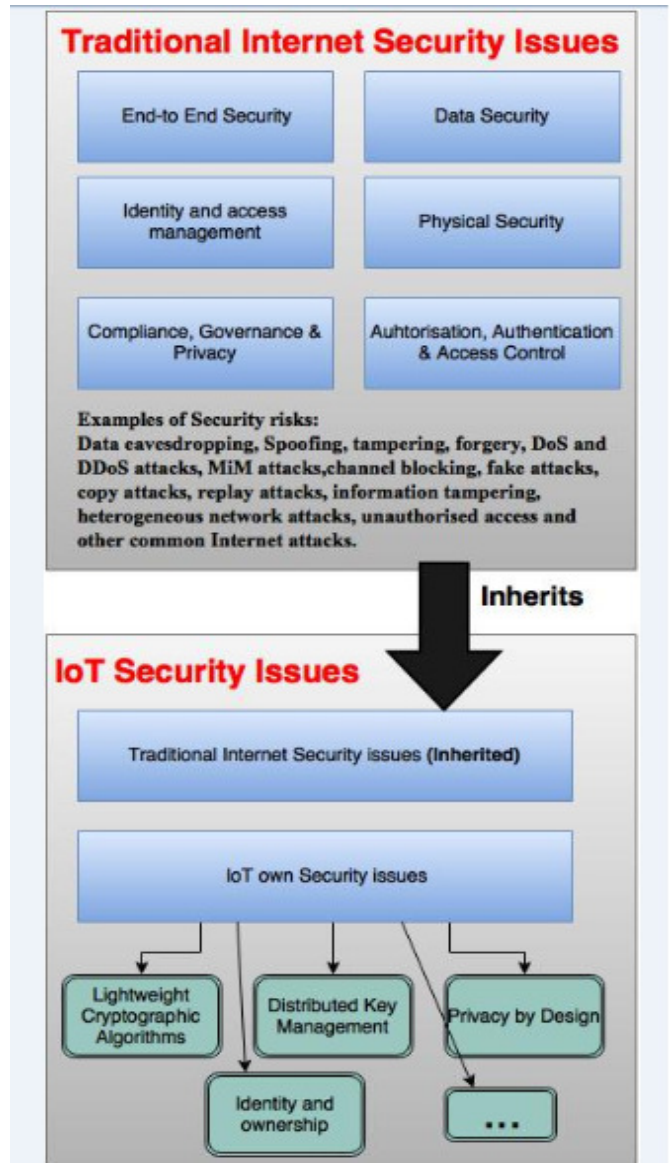

Figure 4- Some IoT security issues

### 4.1.2. Data Security

Data security involves the protection of data during communications and storages. In [26], data security is defined as the process of protecting data from destructive forces or from unauthorized access. Data security, also referred to as information security, is vital to the IoT security. Data security in the IoT is also associated with safety. Usually, the impact of data security breaches on the human life remained within the scope of hacking the personal information of an individual or getting unauthorised access to sensitive information such as financial data. However, data security breaches in the IoT could pose a serious threat to humans' safety. For instance, the accidental intrusion or malicious access that could interfere or interrupt the operations of a driverless car or a heart pacemaker will threaten the user's life. Security breaches in an IoT forest fire detection system could lead to catastrophic results as well.

### 4.1.3. Identity and Access Management

Identity theft, forgery, and masquerading among other security attacks are some of the security issues challenging the protection of identity in the IoT. As we have seen in section 3.2, things identifiers remains an unresolved issue in the IoT. How things are identified, represented, searched, and accessed in the IoT is still unknown. This is indeed make things vulnerable to many identity attacks. For instance, a device in the IoT could uses a fake identity to gain unauthorized access to services provided by another IoT device. This type of attack is known as the masquerade attack. Typically, computer devices employ secure mechanisms that rely on complex algorithms in detecting suspicious access to data and detect imposters. The IoT is vulnerable to several identity attacks including Spoofing, Masquerade, MiM and Smurf attacks as well. Thus, several existent traditional security solutions need to be studied and examined to determine their feasibility and applicability in the IoT.

### 4.1.4. Compliance

Complying with government laws and industry regulations plays an important role in preserving the security of IoT systems. Things in the IoT need to adhere to several data protection laws and privacy acts. Privacy in the IoT requires special considerations as well. This is because the IoT is built around autonomous communications between things. Therefore, there is a need to ensure privacy at all time. Initially, privacy requirements can be summarized to three key concepts:

1. User consent: the user needs to be able to provide an informed consent on the usage of their data
2. Freedom of choice: the user should have the freedom of opting in and out from being involved or being part of a communication
3. Anonymity: the user has the right to remain anonymous when obtaining services that do not require identity verification or the like.

### 4.1.5. Access Control

Access control is the process of granting, limiting or restricting access to a resource. It regulates who or what can view or use resources. Role based access control (RBAC) is an example of a widely used access control model. Access control in the IoT is discussed in more details in a subsequent section. Henceforth, as illustrated in Figure 4, in addition to the security issues inherited from the traditional Internet, the IoT has a set of specific and new security problems. These security issues are derived from the dynamic IoT network structure, the different type of communications involved, and the low-cost characteristic of the IoT devices among other factors which are exclusively associated with the IoT. Significantly, these security issues cannot be solved with traditional Internet security solutions. This is due to the fact that IoT communications' architectures differ from those of the traditional Internet. As the IoT evolves and becomes more complex, the security issues increase in complexity as well. This increase in complexity can be attributed to two fundamental IoT factors: low-cost and heterogeneity.

With regard to low-cost, some IoT devices should be available at relatively low prices. The low-cost of things is a significant factor that drives the support for large-scale deployment of things in the IoT. However, this low-cost requirement dictates that things are mostly resource constrained. This translates into devices with lower computational capabilities, limited amount of memory and power supply. This is, in fact, constitute an obstacle for the application of many traditional cryptographic-based solutions. Given that traditional public-key infrastructures cannot accommodate the IoT [27]. For instance, the adoption of many traditional and basic Internet security solutions such as PKI and CA increases the cost of the IoT devices.

As of heterogeneity, the diversity of devices and communications in the IoT produce many new security challenges as well. For instance, the integration of WSNs into the Internet, as part of the IoT, creates new security problems. These security problems are derived from the process of connecting a sensor node with an Internet device. For example, low-cost and constrained devices that use low-power communication technologies, such as ZigBee or IEEE 802.11ah, need to establish a secure communication channel with more capable devices such as a smartphone. Thus, securing this communication channel requires the use of adequate cryptographic and key management solutions without consuming a lot of bandwidth and energy. This is in addition to the need to employ security protocols which securely connect these devices to the Internet. Add another essential security requirement, such as the need to authorize the devices involved in the communications, and the degree of achieving security for this simple communication scenario increases in complexity. In [28], the authors analyzed some of the major security issues challenging the IoT, which lead to the proposition of some key technologies based on access control and user authentication. Similar works proposed solutions based on the key management architecture such as in [29]. While in [30], protecting the privacy of information in the IoT was the focus of the proposed architecture. However, these contributions are still in design stage as such the details of the protocols and their implementations are missing.

Consequently, given that the IoT merges the traditional Internet, WSNs, PANs, Low-Power wireless networks and many other evolving networks together, the current Internet security solutions can be used to provide some security for the IoT. However, conventional Internet security technologies cannot provide a complete security solution for the IoT. The heterogeneity of devices and the multi-networks integration characteristic of the IoT in addition to the limited capabilities and low-cost requirement of some IoT devices create a new set of IoT own security problems. We will discusses in more details some of the IoT new security issues.

### 4.1.6. Physical and DoS Security Risks

Traditionally, network equipment require the protection against physical attacks and against unauthorized accesses e.g. the storage of routers in secure cabinets. In the IoT, many IoT devices require similar protection measures against physical or unauthorized attacks. Hence, strengthening the physical security of things is essential in many IoT applications. The IoT is vulnerable to the Denial of Service (DoS) attack as well. Typically, a DoS attack floods a given server with false requests for services. Thus, it prevents legitimate requesters from accessing the server's services [31]. It attempts to exhaust the computational resources of the server. The IoT vulnerability to the DoS attack is not only limited to things which connect to the Internet directly, but also extend to WSNs. We have seen in Section 3.2 that nodes in a WSN connect to the Internet eventually, despite the topology used in the network. Therefore, WSNs cannot escape DoS attacks [32]. Additionally, the heterogeneous nature and complexity of communications envisioned in the IoT, makes the IoT vulnerable to the distributed denial of service (DDoS) attack. A DDoS is a DoS attack made by multiple agents in the network and from various locations [33]. Therefore, disruptive attacks such as the DoS and DDoS attacks are a serious potential risk to the IoT. Many IoT devices have limited processing capabilities and memory constraints. Therefore, DDoS attacks can easily exhaust their resources. Also, in things- to-things communications, DoS attacks can prove to be difficult to notice before the disruption of the service which could generally be attributed to battery exhaustion [34].

As a DoS countermeasure, many protocols such as DTLS, IKEv2, HIP, and Diet HIP, verifies the address of the initiating host before responding to requests [34]. Other DoS resistance methods rely on clustering techniques for detecting DoS attacks in WSNs. For example, the work in [35], utilizes a hierarchical clustering technique which detect abnormal behavior inside a cluster by analyzing the traffic on the network using an elected node. Other solutions are centered on the design of intrusion detection system (IDS) such as the one developed specifically to work with WSNs in [36]. SVELTE [37] is designed for 6Lowpan networks and aims to protect against routing attacks as well [37]. For further readings on the defense mechanisms available against DDoS attacks, the reader is referred to [38]. However, it should be noted that the practical implementations and performance of these methods in the IoT are yet to be explored and evaluated.

## 4.2 IoT New Security Challenges

Authorization, authentication, integrity, trust and confidentiality are all fundamental aspects of security which need to be addressed in the IoT. However, securing the IoT communications is challenged by the lack of a shared infrastructure and common security standards. This is, in fact, poses many new security issues which are by far more complicated than those found in any existing networking systems. Table 2 summarizes some of these security requirements:

Table 2- IoT Security Requirements

| | |
|---|---|
| **Authorization** | For smart IoT devices, this requirement can be satisfied using traditional authorization techniques. For constrained devices and in low-power wireless networks, such as ZigBee IP, unauthorized access to the IoT devices should be blocked at the coordinator. That's unauthorized requests should not even be routed to the IoT devices as this may exhaust their energy. |
| **Authentication** | Authentication simply means verifying that "you are who you are claiming to be". This is usually done using a username and password based authentication system. However, this system is not secure enough. Passwords usually require frequent changing and it cannot be used with unattended devices. Also, the Secure Sockets Layer protocol (SSL) is used for authentication. (Mainly, a web browser authenticates web sites using SSL). Authentication also include the process of authenticating both the sender and receiver where they are able to verify the origin of the exchanged messages. This is a complicated security requirement in the IoT. This is because things might not necessarily have IP addresses. |
| **Integrity and Freshness** | Message integrity is about ensuring a message hasn't been altered. This is extremely important in the IoT as many applications rely on information supplied by things to change the statuses of other things. Freshness is also vital for ensuring that no older messages are replayed. |
| **Confidentiality** | Protecting personal and sensitive data from being accessed by unauthorized entities in lightweight devices in the IoT is a challenging task. |

### 4.2.1. Access Control in the IoT

Role Based Access Control (RBAC) is a widely adopted access control approach that restricts access to systems or resources to authorized users. RBAC is based on the concept of assigning permissions to roles. For example, within a healthcare organization, roles are created for various job functions such as a nurse and a doctor. The permissions to perform certain operations in a system are assigned to specific roles. Medical staff (or other system users) are assigned to a particular roles. Through these role assignments, the staff acquire the permissions to perform particular system functions. Since users are not assigned permissions directly, but only acquire them through their roles, management of individual users' privileges becomes a matter of simply assigning an appropriate role to the user. This simplifies common operations, such as adding a user or changing a user's role within a department.

However, in the IoT, things and users might require to access data anytime, anywhere, from various types of devices, including mobile things (things on the move). Therefore, the IoT poses new challenges to access control. In fact, the low power requirements of things, limited bandwidth, heterogeneity of communications and the large scale of devices in the IoT create a unique set of access control requirements. Thus, traditional access control systems in their current status, such as RBAC, might not work efficiently in the IoT. Classically, the advantage of using RBAC in a system is the ability of easily adding access rights to a user, as long as it uses existing roles. In the IoT, as the number of connected devices and networks grow, the number of users and devices requesting access to data and services grow as well. Therefore, an IoT system that comprises thousands of devices would perhaps ends up with thousands of roles and permissions that need to be maintained. Therefore, the challenge of managing access control to thousands of devices in the IoT will be simply transformed into a complex task that requires the management of a vast number of roles. This is known as "role explosion" and it is a major drawback of RBAC. This is because adding new roles to a system require a system-wide update. Generally, RBAC systems are implemented in a centralized architecture whereas an access control server assigns roles to users and grants them access to resources. Thus, updates at the system level are even harder to implement in the IoT given its distributed architecture.

Attribute based access control (ABAC) is an access control model that abstract identity, role, and resources information of the traditional access control into entity attributes [39]. ABAC grants accesses to services based on the attributes possessed by the requester [40]. Therefore, unlike RBAC, the ABAC model uses attributes to describe requesters and the requested services. The associated attributes of each entity can be defined according to the system needs [41]. Relying on attributes provides a much more fine-grained access control approach. As such access control strategies can be designed to use not only the requester' attributes but also other contextual data e.g. the location of the requester. Therefore, rather than grating or denying access to a resource based on the role of a user, ABAC combines together the user's attributes along with other contextual information which provide a way of dynamically generating context aware decisions for requests [42]. This makes ABAC suitable for adoption in systems that require fine grained access control such as the IoT. Additionally, ABAC better adapts to the access control requirements in the IoT including the dynamic expansion of large scale users and things. The advantages of using ABAC in the IoT can be summarized to as follows:

- Data minimization: ABAC uses attributes to identify requesters. Significantly, resources are also represented using an attributes schema. Therefore, ABAC can be used to provide access only to the personal data required for the provision of a request to a service. This concept is known as data minimization
- Privacy enhancement: ABAC supports data minimization. This allows an organization to better design privacy statements where only the required data are specified.

- Policy driven decisions: Access decisions to resources are managed via policies rather than by individuals.
- Dynamic: Access decisions are made during runtime since they use context information as well. Example: let's assume there is exist a policy that grants access to a specific resource only to requesters located in Sydney. Therefore, the system during runtime will need to check the location of the requester before making an access decision.
- Flexibility: Access decisions are associated with access policies rather roles. This allows the system to be more flexible in designing policies according to needs. This is because permissions are no longer bound to roles as in the RBAC. Instead, RBAC allows policies' customizations that leverage accesses to resources. This feature of RBAC enables owners or administrators to apply access control policies on things without the need to acquire a prior knowledge on the requester(s). For example, when a new device joins a system, there will be no need for the rules or policies to be modified. As long as the new device is assigned the attributes necessary for accessing the required resources. Therefore, the ability of accommodating external entities is one of the primary benefits of using ABAC in the IoT. ABAC rules can evaluate attributes of a subject and resources that are not necessary inventoried in the authorization system as well.
- Fine grained: ABAC rules can be fine-grained and contextual.

Consequently, ABAC is a promising access control technology for the IoT. Many researches have started to explore the benefits of using ABAC for managing access control to things and resources in the IoT. For instance the work in [39], proposes an authentication method based on ABAC for the IoT. In [43], the authors highlight the importance of using ABAC in the IoT and Big Data. They provide application examples of ABAC. Also, the study discusses some potential IoT applications where ABAC can be used to provide access control. For example, airlines, financial sectors, exporters, hospitals, among other organizations are described as possible IoT applications that could benefit from the use ABAC in managing access control. For instance in the health sector, the study points out that from a patient perspective ABAC can be used to effectively collect users' consent for the management of patients' EHRs. Moreover, an access control solution based on ABAC policy can be used to provide patients with a granular capability allowing them to manage their personal information [43].

## 5. PRIVACY CHALLENGES

The IoT highly distributed nature of technologies, such as embedded devices in public areas, create weak links that malicious entities can exploit and can as well open the door for a mass surveillance, tracing, tracking, and profiling of the users' movements and activities [44]. Nowadays, the proliferation of mobile devices, GPS systems and other evolving technologies into our lives has introduced several privacy threats. For instance, the study in [45] showed that a driver's home location can be inferred from the GPS data collected from his vehicle even if the location information was anonymized. It further shows that the reconstruction of an individual's route could provide a detailed movement profile that allows for inferences. For example, recurring visits to a medical clinic could indicate illness and visits to activist organizations could hint at political opinion. Other studies such as [46] reported some privacy incidents from the use of mobile applications on the Android, Blackberry, iPhone, and Windows Phone platforms. Thus, it is most likely that privacy incidents will grow rapidly with the increase in penetration of the IoT in our daily life.

Therefore, privacy is one of the major implications as the Internet of Things develops. Privacy no longer means anonymity in the IoT. Profiling and data mining within any IoT scenario can form a potential harm to individuals due to the automatic process of data collection, their storage and the way personal data can be easily shared and analyzed. One of the promising features of the IoT is the ability of devices to observe and sense their environments. Thus, attackers may configure devices to join a given IoT system or network for the purpose of miscellaneously collecting information about the system environment and the user. The most serious concern with things in the IoT is their ability to initiate, by themselves, exchange of information with each other's. Smart devices such as smart home appliances, smart cars and others that log data about their environment, e.g., their locations, constitute a source of risks and vulnerabilities, with regard to privacy, to their owners. If these devices are connected together, as envisioned in the IoT, and these logs are shared among IoT applications, then there is an increased risk of personal information leakage which threaten the users' privacy.

# 6. CONCLUDING REMARKS

In the early days of the Internet which was basically centered on computers, a network of networks was the term used to define the Internet. In the IoT, it seems everything is going to be connected, pants, shoes, shirts, fridges, glasses, washing machines, plants, dogs, cars, airplanes, cities, you name it. Yet, the term network of networks can still be used to define the IoT. However, what's new is that connected networks are no longer limited to IP connected devices/networks in the fashion that we know today. Instead, there are islands of networks connecting using various network technologies. This paper reviewed some of the major challenges facing the IoT. Today, the majority of IoT devices are connected to a mobile phone application. To get anything done, the user has to administer an array of applications in addition to jumping from an application to another in order to control what should be a smart device. We are at risk of creating remote islands of IoT technologies instead of achieving a true vision of IoT. The IoT has the potential of shaping the way we consume energy, improve resources efficiency such as food and water, and support assisted living, access to healthcare, and so much more by connecting different applications together. Therefore, for the realization of a true vision of the IoT, major challenges such as achieving interoperability between the various IoT enabling technologies and devices were identified in this research. Additionally, the main challenge is not only in simply building an IoT system that connects various IoT devices together, but in maintaining scalable, private, secure and trustworthy operations on the IoT. Consequently, it is concluded that there is a need to accommodate the differences in technologies across the various areas of the IoT.

For the future advancement of the IoT, it is therefore imperative to develop a multifaceted technology approach to IoT security, interoperability, management and privacy. The internetworking mechanisms of things, WSNs and traditional computer devices in the IoT are vital with respect to standardizing the communications on the Internet. It is also crucial to have lightweight, scalable and adaptive security solutions in place to secure the users' information and preserve their privacy in the IoT.

## ACKNOWLEDGEMENTS

This research is supported by the International Postgraduate Research Scholarship (IPRS) and the Australian Postgraduate Award (APA).

**AUTHORS**

Mahmoud Elkhodr is with the School of Computing, Engineering and Mathematics at Western Sydney University (Western), Australia. He has been awarded the International Postgraduate Research Scholarship (IPRS) and Australian Postgraduate Award (APA) in 2012-2015. Mahmoud has been awarded the High Achieving Graduate Award in 2011 as well. His research interests include: Internet of Things, e-health, Human Computer-Interactions, Security and Privacy.

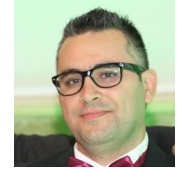

Dr. Seyed Shahrestani completed his PhD degree in Electrical and Information Engineering at the University of Sydney. He joined Western Sydney University (Western) in 1999, where he is currently a Senior Lecturer. He is also the head of the Networking, Security and Cloud Research (NSCR) group at Western. His main teaching and research interests include: computer networking, management and security of networked systems, analysis, control and management of complex systems, artificial intelligence applications, and health ICT. He is also highly active in higher degree research training supervision, with successful results.

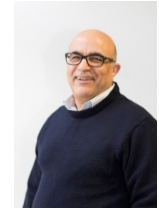

Dr. Hon Cheung graduated from The University of Western Australia in 1984 with First Class Honours in Electrical Engineering. He received his PhD degree from the same university in 1988. He was a lecturer in the Department of Electronic Engineering, Hong Kong Polytechnic from 1988 to 1990. From 1990 to 1999, he was a lecturer in Computer Engineering at Edith Cowan University, Western Australia. He has been a senior lecturer in Computing at Western Sydney University since 2000. Dr Cheung has research experience in a number of areas, including conventional methods in artificial intelligence, fuzzy sets, artificial neural networks, digital signal processing, image processing, network security and forensics, and communications and networking. In the area of teaching, Dr Cheung has experience in development and delivery of a relative large number of subjects in computer science, electrical and electronic engineering, computer engineering and networking.

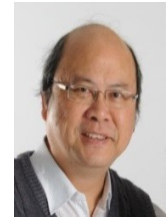